\documentstyle[12pt,
epsfig]{article}
\begin{document}
%
\relax
\renewcommand{\theequation}{\arabic{section}.\arabic{equation}}
\def\be{\begin{equation}}
\def\ee{\end{equation}}
\def\bs{\begin{subequations}}
\def\es{\end{subequations}}

\def\calm{{\cal M}}
\def\lx{\lambda}
\def\ex{\epsilon}
\def\Lx{\Lambda}

\newcommand{\tl}{\tilde t}
\newcommand{\ttt}{\tilde T}
\newcommand{\rhot}{\tilde \rho}
\newcommand{\ptt}{\tilde p}
\newcommand{\drho}{\delta \rho}
\newcommand{\drhot}{\delta {\tilde \rho}}
\newcommand{\dchi}{\delta \chi}

\newcommand{\A}{A}
\newcommand{\B}{B}
\newcommand{\mmu}{\mu}
\newcommand{\mnu}{\nu}
\newcommand{\ii}{i}
\newcommand{\jj}{j}
\newcommand{\jl}{[}
\newcommand{\jr}{]}
\newcommand{\ml}{\sharp}
\newcommand{\mr}{\sharp}

\newcommand{\da}{\dot{a}}
\newcommand{\db}{\dot{b}}
\newcommand{\dn}{\dot{n}}
\newcommand{\dda}{\ddot{a}}
\newcommand{\ddb}{\ddot{b}}
\newcommand{\ddn}{\ddot{n}}
\newcommand{\pa}{a^{\prime}}
\newcommand{\pn}{n^{\prime}}
\newcommand{\ppa}{a^{\prime \prime}}
\newcommand{\ppb}{b^{\prime \prime}}
\newcommand{\ppn}{n^{\prime \prime}}
\newcommand{\fda}{\frac{\da}{a}}
\newcommand{\fdb}{\frac{\db}{b}}
\newcommand{\fdn}{\frac{\dn}{n}}
\newcommand{\fdda}{\frac{\dda}{a}}
\newcommand{\fddb}{\frac{\ddb}{b}}
\newcommand{\fddn}{\frac{\ddn}{n}}
\newcommand{\fpa}{\frac{\pa}{a}}
\newcommand{\fpb}{\frac{\pb}{b}}
\newcommand{\fpn}{\frac{\pn}{n}}
\newcommand{\fppa}{\frac{\ppa}{a}}
\newcommand{\fppb}{\frac{\ppb}{b}}
\newcommand{\fppn}{\frac{\ppn}{n}}

\newcommand{\pt}{\tilde{p}}
\newcommand{\rhb}{\bar{\rho}}
\newcommand{\pb}{\bar{p}}
\newcommand{\pbb}{\bar{\rm p}}
\newcommand{\rht}{\tilde{\rho}}
\newcommand{\kt}{\tilde{k}}
\newcommand{\kb}{\bar{k}}
\newcommand{\wt}{\tilde{w}}

\newcommand{\dA}{\dot{A_0}}
\newcommand{\dB}{\dot{B_0}}
\newcommand{\fdA}{\frac{\dA}{A_0}}
\newcommand{\fdB}{\frac{\dB}{B_0}}

\def\be{\begin{equation}}
\def\ee{\end{equation}}
\def\bs{\begin{subequations}}
\def\es{\end{subequations}}
\newcommand{\een}{\end{subequations}}
\newcommand{\ben}{\begin{subequations}}
\newcommand{\beq}{\begin{eqalignno}}
\newcommand{\eeq}{\end{eqalignno}}
\def \lta {\mathrel{\vcenter
     {\hbox{$<$}\nointerlineskip\hbox{$\sim$}}}}
\def \gta {\mathrel{\vcenter
     {\hbox{$>$}\nointerlineskip\hbox{$\sim$}}}}

\def\g{\gamma}
\def\mpl{M_{\rm Pl}}
\def\ms{M_{\rm s}}
\def\ls{l_{\rm s}}
\def\l{\lambda}
\def\m{\mu}
\def\n{\nu}
\def\a{\alpha}
\def\b{\beta}
\def\gs{g_{\rm s}}
\def\d{\partial}
\def\co{{\cal O}}
\def\sp{\;\;\;,\;\;\;}
\def\r{\rho}
\def\dr{\dot r}
\def\e{\epsilon}
\newcommand{\NPB}[3]{\emph{ Nucl.~Phys.} \textbf{B#1} (#2) #3}   
\newcommand{\PLB}[3]{\emph{ Phys.~Lett.} \textbf{B#1} (#2) #3}   
\newcommand{\ttbs}{\char'134}           
\newcommand\fverb{\setbox\pippobox=\hbox\bgroup\verb}
\newcommand\fverbdo{\egroup\medskip\noindent%
                        \fbox{\unhbox\pippobox}\ }
\newcommand\fverbit{\egroup\item[\fbox{\unhbox\pippobox}]}
\newbox\pippobox
\def\tr{\tilde\rho}
\def\lb{w}
\def\bbox{\nabla^2}
\def\mt{{\tilde m}}
\def\rct{{\tilde r}_c}
\def \lta {\mathrel{\vcenter
     {\hbox{$<$}\nointerlineskip\hbox{$\sim$}}}}
\def \gta {\mathrel{\vcenter
     {\hbox{$>$}\nointerlineskip\hbox{$\sim$}}}}
\noindent
\begin{flushright}
June 2004
\\
\end{flushright} 
\vspace{3cm}
\begin{center}
{ \Large \bf
Brane Cosmology with Matter in the Bulk (II)
\\
} 
\vspace{1.5cm}
{\Large 
N. Tetradis 
} 
\\
\vspace{0.5cm}
{\it
Department of Physics, University of Athens,
Zographou 157 71, Greece
} 
\\
\vspace{3cm}
\abstract{
We derive exact solutions of the Einstein equations in the context of the
Randall-Sundrum model with matter on the brane and in the bulk. We present 
two models in which the brane moves within a time-dependent bulk.
We study the cosmological evolution on the brane. Our solutions display
novel behaviour, such as an 
expansion driven only by the bulk matter and the appearance
of a ``mirage'' dust component on the brane.
} 
\end{center}

\newpage

\section{Introduction}
\setcounter{equation}{0}

The Randall-Sundrum model \cite{rs} provides the necessary framework
in which a consistent 
Friedmann-Robertson-Walker cosmology can 
emerge in a space-time with more than three spatial dimensions 
\cite{binetruy,csaki,kraus}.
The Universe is identified with a four-dimensional hypersurface
(a 3-brane) in a five-dimensional bulk with negative cosmological
constant (AdS space). The geometry is non-trivial (warped)
along the fourth spatial dimension, so that an effective localization of
low-energy gravity takes place near the brane. The matter is assumed to be
concentrated only on the brane.
For low matter densities the cosmological evolution on the brane
becomes identical to the standard
Friedmann-Robertson-Walker cosmology. 
Several aspects of the cosmological evolution 
have been the subject of a large number of studies. (For recent reviews,
with extensive lists of references, see ref. \cite{brax}.)

If the bulk contains some matter component in addition to the negative
cosmological constant the brane evolution is modified. For example, it is
possible to have energy exchange between the brane and the bulk. Also the
presence of a fluid in the bulk can alter the expansion on the brane. 
For an observer on the brane the 
modification can be attributed to ``mirage'' matter components 
\cite{brax}--\cite{tet1}.
In the presence of bulk matter, the cosmological evolution on the brane
is not autonomous, and the explicit knowledge of the bulk energy-momentum
tensor is necessary. As a result, certain assumptions often
need to be made for its form \cite{vandebruck}--\cite{tet2}. 
Previous studies have considered a bulk field \cite{dilaton}, or
radiation emitted by the brane into the bulk \cite{vaidya}.
Also an attempt has been made to incorporate thermal effects in the bulk
\cite{chamblin,rothstein}.

In the case of an empty bulk,
the brane evolution can be discussed either in a coordinate system
(system A) in which the brane is located at a fixed value of the fourth spatial
coordinate and the bulk is time-dependent \cite{binetruy,csaki}, or
in a different coordinate system (system B) 
in which the bulk is static and the brane
is moving \cite{kraus}. In the latter case, the bulk metric is 
five-dimensional Schwarzschild-AdS \cite{birm}. The two points of view are
equivalent \cite{gregory}. 

In ref. \cite{pantelis} the interpolation between the two coordinate 
systems was employed in order to construct a new example of brane evolution
in a bulk that contains an ideal fluid. The bulk was assumed to be static in
the coordinate 
system B. A solution was found that is the generalization in an AdS 
background of the well known solutions for the interior of stellar objects.
In the coordinate system A the cosmological evolution on the brane 
is affected by ``mirage'' components arising from the presence of the 
bulk fluid. Moreover, there is energy exchange between the bulk and the 
brane.

In this work we present two different examples of brane evolution.
The new ingredient is that the bulk is time-dependent even in the 
coordinate system B. The brane evolution displays new features, such as
an expansion driven by the bulk instead of the brane matter, or a ``mirage''
component that has the equation of state of dust.

We consider an action of the form
\be
S=\int d^5x~ \sqrt{-g} \left( M^3 R +\Lambda +{\cal L}_B^{mat}\right)
+\int d^4 x\sqrt{-\gamma} \,\left( -V+{\cal L}_b^{mat} \right), 
\label{001}
\ee
where $R$ is the curvature scalar of the five-dimensional metric 
$g_{AB}, A,B=0,1,2,3,4$,
$-\Lambda$ the bulk cosmological constant ($\Lx >0$),
${\gamma}_{ab}$, $a,b=0,1,2,3$,
the induced metric on the 3-brane, and  
$V$ the brane tension.
In our discussion of the bulk metric we shall 
describe the bulk matter in terms of the energy-momentum tensor 
$T^{A}_{~B}$ of
a fluid. The brane matter will be described in a similar way.
Energy flow can appear
along the fourth spatial dimension, described by
the off-diagonal element $T^0_{~4}$. 
The presence of a flow requires an appropriate interaction between the
brane and bulk matter. We implicitly assume that 
the Lagrangian ${\cal L}_b^{mat}$ contains the necessary terms for the 
energy exchange between the bulk and the brane that is consistent with
the form of $T^0_{~4}$ in our solutions.

\section{Model I}
\setcounter{equation}{0}

\subsection{The bulk}

We assume that the metric in the bulk can be put in the form
\begin{equation}
ds^{2}=-n^{2}(r) dt^{2}+c^2(t) r^{2}d\Omega_k^2
+b^{2}(r)dr^{2},
\label{metric}
\end{equation}
where $d\Omega^2_k$ is the metric in a
maximally symmetric three-dimensional space. 
We use $k=-1,0,1$ to parametrize the spatial curvature.
The non-zero components of the five-dimensional Einstein tensor are
\begin{eqnarray}
{G}^{\,0}_{~0} &=&  \frac{3}{b^2} \frac{1}{r}
\left( \frac{1}{r}-\frac{b_{,r}}{b} \right) 
-\frac{3}{n^2}\left(\frac{c_{,t}}{c} \right)^2
-\frac{3k}{c^2r^2}
\label{ein00} \\
 {G}^{\,i}_{~j} &=& 
\frac{1}{b^2}\left[
\frac{1}{r}\left( \frac{1}{r}+2\frac{n_{,r}}{n}\right) 
-\frac{b_{,r}}{b}\left( \frac{n_{,r}}{n}+2\frac{1}{r}\right)
+\frac{n_{,rr}}{n} \right]
-\frac{1}{n^2}\left[ \left(\frac{c_{,t}}{c} \right)^2
+2\frac{c_{,tt}}{c}\right]
-\frac{k}{c^2r^2}
\nonumber \\
&~&
\label{einij} \\
{G}^{\, 4}_{~4} 
&=& \frac{3}{b^2}\frac{1}{r}\left(\frac{1}{r}+\frac{n_{,r}}{n} \right)
-\frac{3}{n^2}\left[ \left(\frac{c_{,t}}{c} \right)^2
+\frac{c_{,tt}}{c}\right]
-\frac{3k}{c^2r^2},
\label{ein44} \\
{G}^{\, 0}_{~4} 
&=& 
-\frac{3}{n^2}
\left( \frac{n_{,r}}{n}\frac{c_{,t}}{c}- \frac{1}{r}\frac{c_{,t}}{c}
 \right)
\label{ein04} 
\end{eqnarray} 
where the subscripts indicate derivatives with respect to
$r$ and $t$.

The Einstein equations take the usual form 
\begin{equation}
G^{A}_{~B}
= \frac{1}{2 M^3} T^{A}_{~B} \;, 
\label{einstein}
\end{equation}
where $T^A_{~B}$ denotes the total energy-momentum tensor.
Assuming a negative cosmological constant $-\Lx$ and an
anisotropic perfect fluid in the bulk, 
we write the bulk energy-momentum tensor as
\begin{equation}
T^A_{~B}=
{\rm diag}(\Lambda-\rho,\Lambda+p,\Lambda+p,\Lambda+p,\Lambda+{\rm p}).
\label{tmn4}  
\end{equation}

We look for solutions that satisfy the ansatz
\be
n^2(r)=\frac{1}{b^2(r)}=\frac{1}{12M^3}\Lx r^2.
\label{nb} \ee
The Einstein equations (\ref{ein00})--(\ref{ein44})
become
\begin{eqnarray}
\left(\frac{c_{,t}}{c} \right)^2
&=&\frac{1}{6M^3} \rhb(t)-\frac{\kb}{c^2}
\label{ein00a} \\
\left(\frac{c_{,t}}{c} \right)^2
+2\frac{c_{,tt}}{c}
&=&-\frac{1}{2M^3} \pb(t)-\frac{\kb}{c^2}
\label{einija} \\
\left(\frac{c_{,t}}{c} \right)^2
+\frac{c_{,tt}}{c}
&=&-\frac{1}{6M^3} \pbb(t)-\frac{\kb}{c^2},
\label{ein44a} 
\end{eqnarray} 
where $\kb=\Lx k/(12M^3)$. We have assumed that the matter distribution
has the form
\be
\rho(t,r)=\frac{\rhb(t)}{n^2}=\frac{12M^3}{\Lx}\frac{1}{r^2}\,\rhb(t),
\label{matt} \ee 
and similarly for $p(t,r)$ and ${\rm p}(t,r)$.
Eqs. (\ref{einija}), (\ref{ein44a}) can be replaced by 
\begin{eqnarray}
\rhb_{,t}+3\frac{c_{,t}}{c}(\rhb+\pb)&=&0
\label{encons} \\
2{\rm p}-3p+\rho&=&0.
\label{momcons}
\end{eqnarray}
These equations could have been obtained from the conservation of the
energy-momentum tensor. 
The most natural way to satisfy eq. (\ref{momcons})
is to assume that the bulk fluid is anisotropic with an equation of
state defined through $p=w\rho$, ${\rm p}={\rm w} \rho$, with
\be
2{\rm w}-3w+1 =0.
\label{eqstat} \ee
As a result, the evolution in the bulk is described by 
eqs. (\ref{ein00a}), (\ref{encons}), that have a form identical to
that for the standard Friedmann-Robertson-Walker Universe.

\subsection{The brane}

For the discussion of the cosmological evolution on the brane we follow
ref. \cite{kraus}.
First we consider a coordinate system 
(using Gaussian normal coordinates) in which the metric takes the form
\begin{equation}
ds^2=\gamma_{ab}dx^a dx^b+d\eta^2 
=-m^2(\tau,\eta )d\tau^2
+a^2(\tau,\eta )d\Omega_k ^2+d\eta ^2.
\label{sx2.1ex}
\end{equation}
The brane is located at $\eta=0$. In order to keep our
discussion simple we identify the half-space $\eta>0$ with the half-space
$\eta<0$, in complete analogy to ref. \cite{rs}.
We also redefine the time, so as to 
set $m(\tau,\eta=0)=1$.

Through an appropriate coordinate transformation 
\begin{equation}
t=t({\tau},\eta ),\qquad r=r({\tau},\eta )  \label{sx2.4}
\end{equation}
the metric (\ref{metric}) 
can be written in the form of eq. (\ref{sx2.1ex}).
It is obvious that $a({\tau},\eta )=c(t)\, r$. 
Identification of the metrics (\ref{metric}) and (\ref{sx2.1ex}) requires
\begin{eqnarray}
b^2(r)\left( \frac{\partial r}{\partial {\tau}}\right) ^2-n^2(r)\left( 
\frac{\partial t}{\partial {\tau}}\right) ^2&=&-m^2({\tau},\eta)
  \label{sx2.8}
\\
b^2(r)\left( \frac{\partial r}{\partial \eta }\right) ^2-n^2(r)\left( \frac{%
\partial t}{\partial \eta }\right) ^2&=&1  \label{sx2.9}
\\
b^2(r)\,\,\frac{\partial r}{\partial {\tau}}\frac{\partial r}{\partial
\eta }-n^2(r)\,\,\frac{\partial t}{\partial {\tau}}\frac{\partial t}{\partial
\eta }&=&0.  \label{sx2.10}
\end{eqnarray}

We define $R({\tau})=r({\tau},\eta=0)$. 
In the system of coordinates $(t,r)$ of eq. (\ref{metric})
the brane is moving, as it is located at $r=R({\tau})$. 
At the location of the brane, we find 
\begin{eqnarray}
\frac{\partial t}{\partial \tau} &=& 
\frac{1}{n(R)} \left[ b^2(R) \dot{R}^2 +1 \right]^{1/2}
\label{tr1} \\
\frac{\partial t}{\partial \eta} &=& 
-\frac{b(R)}{n(R)} \dot{R}
\label{tr2} \\
\frac{\partial r}{\partial \tau} &=& 
\dot{R}
\label{tr3} \\
\frac{\partial r}{\partial \eta} &=& 
-\frac{1}{b(R)} \left[ b^2(R) \dot{R}^2 +1 \right]^{1/2},
\label{tr4} \end{eqnarray}
where the dot denotes a derivative with respect to proper time.
The negative sign in the r.h.s. of eq. (\ref{tr4}) is a consequence of
our assumption that $r$ decreases on both sides of the brane
\cite{kraus}.

It is possible now to rewrite eqs. (\ref{ein00a}), (\ref{encons})
at the location of the brane using proper-time derivatives.
We find
\begin{eqnarray}
\left( \frac{\dot{c}}{c} \right)^2&=&
 \left( \frac{1}{6M^3} \rho
-\frac{k}{c^2R^2} \right)
\left[1+\frac{12M^3}{\Lx} \left(\frac{\dot{R}}{R}\right)^2\right]
\label{eqone} \\
\dot{\rhb}+3\frac{\dot{c}}{c}(\rhb+\pb)&=&0.
\label{eqtwo} 
\end{eqnarray}
We have denoted
\begin{eqnarray}
c(\tau)&=&c\biggl(t(\tau,\eta=0)\biggr)
\label{ctau} \\
\rho(\tau)&=&\frac{12M^3}{\Lx}\frac{1}{R^2(\tau)}\,\, 
\rhb(\tau)=\frac{12M^3}{\Lx}\frac{1}{R^2(\tau)}\,\, 
\rhb\biggl(t(\tau,\eta=0)\biggr),
\label{rhotau} \end{eqnarray}
in agreement with eq. (\ref{matt}).
Similar notation will be used for $p$ and ${\rm p}$ in the following.
We can rewrite eq. (\ref{eqtwo}) as
\be
\dot{\rho}+2\frac{\dot{R}}{R}\rho+\frac{\dot{c}}{c}(\rho+p)=0.
\label{eqtwoex} \ee

It is straightforward to express the bulk energy-momentum tensor 
at the location of the brane in the coordinate system 
$({\tau},\eta)$. We find
\begin{eqnarray}
T^0_{~0}&=&\Lx-\rho-
b^2(R)\dot{R}^2(\rho+{\rm p} )=\Lx-\rho-
\frac{6M^3}{\Lx} \left(\frac{\dot{R}}{R}\right)^2(\rho+3p )
\label{t00} \\
T^1_{~1}&=&T^2_{~2}=T^3_{~3}
= \Lx +p
\label{t11} \\
T^4_{~4}&=&
\Lx+{\rm p}+b^2(R)\dot{R}^2(\rho+{\rm p})
=\Lx+\frac{1}{2}\left(3 p-\rho \right)  +
\frac{6M^3}{\Lx} \left(\frac{\dot{R}}{R}\right)^2(\rho+3p)
\label{t44} \\
T^0_{~4}
&=&b(R)\dot{R}\left( b^2(R) \dot{R}^2+1 \right)^{1/2}
( \rho+{\rm p})
\nonumber \\
~&=&\left(\frac{3M^3}{\Lx}\right)^{1/2}\frac{\dot{R}}{R}
\left[1+\frac{12M^3}{\Lx} \left(\frac{\dot{R}}{R}\right)^2\right]^{1/2}
( \rho+3p),
\label{t04} 
\end{eqnarray}
where we have made use of eq. (\ref{momcons}).

The equations governing the cosmological evolution on the brane 
can be obtained in various ways. The most straightforward is
to write the Einstein equations in the system $({\tau},\eta)$,
assuming that the brane energy-momentum tensor has the form
\be
\ttt^A_{~B}=
\delta(\eta)\,{\rm diag}(V-\rht,V+p,V+\pt,V+\pt,0),
\label{branet} \ee
with $V$ the brane tension and $\rht$, $\pt$ the energy density and pressure
of a perfect fluid localized on the brane.
The details of the calculation can be found in refs. 
\cite{brax}--\cite{tet2} and
will not be repeated here.
The evolution on the brane is governed by the equations
\begin{eqnarray}
\dot{\rht}+3 H (\rht+\pt)&=&
-2 T^0_{~4}
\label{energ} \\
H^2=\left(\frac{\dot{a}_0^{}}{a_0^{}}\right)^2=
\left( \frac{\dot{R}}{R}+\frac{\dot{c}}{c} \right)^2
&=&\frac{1}{144 M^6}\left(\rht^2+2V\rht\right)
-\frac{k}{a_0^2}+\chi+\phi+\lx
\label{hubble} \\
\dot{\chi}+4H\chi&=&\frac{1}{36M^6}(\rht+V)T^0_{~4}
\label{chi}\\
\dot{\phi}+4H\phi&=&-\frac{1}{3M^3}H\left( T^4_{~4}-\Lx \right),
\label{phi}
\end{eqnarray}
with $a^{}_0(\tau)=a(\tau,\eta=0)=c(\tau)R(\tau)$.
Here $\lx=(V^2/12M^3-\Lx)/12M^3$ is the effective cosmological constant,
and $T^4_{~4}$, $T^0_{~4}$ are given by eqs. (\ref{t44}), (\ref{t04}).

For the particular form of the energy-momentum tensor that we are considering,
it is possible to show that 
\be
\chi+\phi=\frac{2}{\Lx}\frac{\rhb}{R^2}
\label{chiphi} \ee
satisfies eqs. (\ref{chi}), (\ref{phi}).
This permits us to write eq. (\ref{hubble})
in the form
\be
H^2=\left(\frac{\dot{a^{}_0}}{a^{}_0}\right)^2=
\left( \frac{\dot{R}}{R}+\frac{\dot{c}}{c} \right)^2
=\frac{1}{144 M^6}\left(\rht^2+2V\rht\right)
+\frac{1}{6M^3}\rho
-\frac{k}{a_0^2}
+\lx,
\label{hubble2} \ee
with $\rho$ given by eq. (\ref{rhotau}).

An alternative way to derive eq. (\ref{hubble2}) is by considering
the junction conditions for the boundary defined by the brane motion. 
In the coordinates $({\tau},\eta)$ 
of eq. (\ref{sx2.1ex}) the extrinsic curvature of the 
boundary is
\begin{equation}
K_{ab}=\frac{1}{2}\partial_\eta \gamma_{ab}.
\label{extr} \end{equation}
The junction conditions are \cite{israel}
\be
\Delta K_{ab}=K^+_{ab}-K^-_{ab}=-\frac{1}{2M^3}
\left(T_{ab}-\frac{1}{3}T^c_{~c}\gamma_{ab} \right),
\label{junction} \ee
where $\pm$ denote the two sides of the brane, and the
energy-momentum tensor is given by eq. (\ref{branet}) neglecting the
$\delta$-function.
The junction condition for the spatial components leads to 
\be 
\pm  \left( \frac{1}{6M^3} \rho-\frac{k}{c^2R^2} \right)^{1/2}
\left(\frac{12M^3}{\Lx}\right)^{1/2}\frac{\dot{R}}{R}
+\left[\frac{\Lx}{12M^3}+ \left(\frac{\dot{R}}{R}\right)^2\right]^{1/2}
=\frac{1}{12M^3}(\rht+V).
\label{extr1} \ee
The sign of the first term depends on whether $c$ is an increasing or
decreasing function of $\tau$.
If we solve the above equation for $\dot{R}/R$ 
and eq. (\ref{eqone}) for $\dot{c}/{c}$,
we recover eq. (\ref{hubble2}).

We can also use eq. (\ref{eqone}) to rewrite eq. (\ref{extr1})
as
\be
H \frac{\dot{R}}{R}=
\frac{1}{12M^3}\left(\rht+V \right)
\left[\frac{\Lx}{12M^3}+ \left(\frac{\dot{R}}{R}\right)^2\right]^{1/2}
-\frac{\Lx}{12M^3}.
\label{posit} \ee
We are interested in scenaria in which the effective cosmological 
constant $\lx$ is set to zero. This requires the fine-tuning
$V^2/(12M^3)=\Lx$. In this case and for $\rht,V>0$, 
it is clear from the above expression that $H\dot{R}/R >0$.
For an expanding brane Universe with $H>0$, 
and an equation of state
of the bulk matter with $w>-1/3$, ${\rm w}>-1$,
we conclude from eq. (\ref{t04})
that there can be only energy outflow.

\subsection{The cosmological evolution}

The cosmological evolution on the brane is described by
eqs. (\ref{eqone}), (\ref{eqtwoex}), 
(\ref{t04}), (\ref{energ}),
(\ref{hubble2}), while eq. (\ref{extr1}) or (\ref{posit})
can replace one of eqs. 
(\ref{eqone}), (\ref{hubble2}).

The effective Friedmann equation (\ref{hubble2}) for the expansion 
on the brane has several terms: 
The effective cosmological constant 
$\lx$ leads to exponential expansion. In order to
get conventional cosmology we have to set it to zero
through the appropriate fine-tuning of the bulk cosmogical constant and
the brane tension: $V^2=12M^3 \Lx$.
The curvature term has the standard form. 
The brane matter induces a term $\sim \rht^2$, characteristic of the
unconventional early cosmology of the Randall-Sundrum model
\cite{binetruy}. At low energy densities $\rht \ll V$ ,
the standard contribution $\sim \rht$ to the Hubble
parameter dominates. The effective Planck constant is
$\mpl^2=12M^6/V$. 

The bulk matter induces a term linear in the bulk energy density $\rho$. 
This bears no resemble to the 
``mirage'', or ``Weyl'', or ``dark'' radiation term 
\cite{brax,mirage,tet1,hebecker,vaidya} and its variations \cite{pantelis}.
The reason is that the bulk solution of section 2 has been constructed 
in a way that the bulk matter results in an
expansion along the three spatial
dimensions along the brane,
instead of the modification of the warping along the fourth spatial
dimension.

Both the bulk and matter energy densities have unconventional
scaling during the expansion. The bulk energy density $\rho$ obeys 
eq. (\ref{eqtwoex}). It does not scale with a given power of
$a_0^{}$. Instead it falls off as 
$R^{-2}c^{-3(1+w)}=R^{1+3w}a_0^{-3(1+w)}$.
The brane energy density is diluted by the expansion, but 
is also reduced through the outflow from both sides
of the brane. This implies that the energy density of the various components 
of brane matter (dust, radiation etc) does not scale as function
of $R$ with the ``naive'' power $a_0^{-3(1+\wt)}$.
The outflow makes the reduction faster. 
The rate of energy loss, given by eq. (\ref{t04}), is fixed
by the static solution for the bulk that we assumed in section 2. 
For an expanding brane Universe that obeys eq. (\ref{posit}),
this solution is consistent only with $H\dot{R}/R >0$, and therefore 
energy loss by the brane, according to eq. (\ref{t04}).

Generically, the contribution of the bulk energy density $\rho$ 
to the expansion 
becomes dominant over that of the brane energy density $\rht$
at late times. 
The reason is the enhancement factor $R^{1+3w}$ that multiplies the 
conventional scaling $a^{-3(1+w)}_0$ of $\rho$.
Moreover, the brane energy density is reduced faster than the
conventional scaling $a^{-3(1+\wt)}_0$ because of the outflow. 
As $T^0_{~4}$ is proportional to $\rho$ in our
model, we find that generically $\rht$ becomes zero at a finite time.
Beyond this point the consistent solution is $\rht=0$, $\dot{R}/R=0$,
$H=\dot{c}/{c}$.
The cosmological expansion on the brane is driven only 
by the bulk energy density.
The brane remains stationary at a certain point along the fourth
spatial dimension $r=R(\tau)=$constant. 

An interesting feature concerns the possibility of having periods of
accelerated expansion during the evolution. If the bulk energy density 
is assumed to dominate the expansion in eq. (\ref{hubble2}), the 
acceleration parameter can be calculated to be
\be
Q=\frac{1}{H^2} \frac{\ddot{a}^{}_0}{a^{}_0}=
-\frac{1}{H}\, \frac{\dot{c}}{c}\,\frac{1+3w}{2}.
\label{accel} \ee
In models with $\dot{c}/c<0$, $H>0$, we expect acceleration during
this era. As we mentioned above, the brane energy density 
approaches zero shortly after the point that the bulk energy density
dominates the expansion. As a result $H=\dot{R}/{R}+\dot{c}/{c}$ first
becomes zero and subsequently negative following $\dot{c}/{c}$.
This indicates that the periods of accelerated expansion do not last long.

As a final remark we point out that the anisotropic bulk fluid that satisfies
eq. (\ref{eqstat}) can have a simple physical interpretation. For example,
we could have $w=1/3$, w$=0$. The fluid behaves as radiation in the directions
parallel to the brane, but is pressureless along the fourth spatial
dimension.

\section{Model II}
\setcounter{equation}{0}

\subsection{The bulk}

We consider a bulk metric of the form
\begin{equation}
ds^{2}=-n^{2}(u,r)\, du^{2}+2\ex\,du \,dr+ r^{2}d\Omega_k^2,
\label{metric2}
\end{equation}
where 
\be
n^2(u,r)=\frac{1}{12M^3}\Lx r^2+k-\frac{1}{6\pi^2M^3} \frac{\calm(u,r)}{r^3} 
\label{ns} \ee
and $d\Omega^2_k$ is the metric in a
maximally symmetric three-dimensional space ($k=-1,0,1$).
This is a generalized Vaidya metric \cite{wang}, similar to those
used in the discussion of the radiation of bulk gravitons by the 
brane \cite{hebecker,vaidya,chamblin}. 
The parameter $\ex$ takes the
values $\ex=\pm 1$.
In studies of graviton emission it is usually assumed
that $\calm=\calm(u)$ and $\ex=1$. 
Our discussion is more general, as it allows for 
an additional dependence of $\calm$ on $r$. It can also account for
energy absorption by the brane when $\ex=-1$.

The energy-momentum tensor
that satisfies the Einstein equations is
\begin{eqnarray}
{T}^{0}_{~0} = {T}^{4}_{~4} &=& \Lx -\frac{1}{2\pi^2} \frac{\calm_{,r}}{r^3}
\label{t002} \\
{T}^{1}_{~1} = {T}^{2}_{~2} = {T}^{3}_{~3} 
&=& \Lx -\frac{1}{6\pi^2} \frac{\calm_{,rr}}{r^2}
\label{t112} \\
{T}^{0}_{~4} &=& \frac{1}{2\pi^2} \frac{\calm_{,u}}{r^3},
\label{t042} 
\end{eqnarray} 
where the subscripts indicate derivatives with respect to
$r$ and $u$. The matter component that is added to the cosmological
constant satisfies the
various energy conditions 
if one requires that 
$\ex \calm_{,u} \geq 0$, $\calm_{,r}\geq 0$,
$\calm_{,rr}\leq 0$, $\calm_{,r}\geq -r\calm_{,rr}/3$ \cite{wang}.

\subsection{The brane}

For the discussion of the cosmological evolution on the brane we proceed
similarly to the previous section.
We consider a coordinate system 
in which the metric takes the form
\begin{equation}
ds^2=\gamma_{ab}dx^a dx^b+d\eta^2 
=-m^2(\tau,\eta )d\tau^2
+a^2(\tau,\eta )d\Omega_k ^2+d\eta ^2.
\label{sx2.1ex2}
\end{equation}
The brane is located at $\eta=0$, while we
identify the half-space $\eta>0$ with the half-space
$\eta<0$. We also redefine the time, so as to 
set $m(\tau,\eta=0)=1$.

Through an appropriate coordinate transformation 
\begin{equation}
u=u({\tau},\eta ),\qquad r=r({\tau},\eta )  \label{sx2.42}
\end{equation}
the metric (\ref{metric2})
can be written in the form of eq. (\ref{sx2.1ex2}). 
It is obvious that $a({\tau},\eta )=r$. 
Identification of the metrics (\ref{metric2}) and (\ref{sx2.1ex2}) requires
\begin{eqnarray}
-n^2(u,r)\left( \frac{\partial u}{\partial {\tau}}\right) ^2
+2\ex \, \frac{\partial u}{\partial {\tau}}\frac{\partial r}{\partial {\tau}}
&=&-m^2({\tau},\eta)
\label{sx2.82} \\
-n^2(u,r)\left( \frac{\partial u}{\partial {\eta}}\right) ^2
+2\ex \, \frac{\partial u}{\partial {\eta}}\frac{\partial r}{\partial {\eta}}
&=&1
\label{sx2.92} \\
-n^2(u,r) \, 
\frac{\partial u}{\partial {\tau}}\frac{\partial u}{\partial {\eta}}
+\ex \, \frac{\partial u}{\partial {\tau}}\frac{\partial r}{\partial {\eta}}
+\ex \, \frac{\partial u}{\partial {\eta}}\frac{\partial r}{\partial {\tau}}
&=&0.  
\label{sx2.102}
\end{eqnarray}

We define $R({\tau})=r({\tau},\eta=0)$. 
In the system of coordinates $(u,r)$ of eq. (\ref{metric2})
the brane is moving, as it is located at $r=R({\tau})$. 
At the location of the brane, we find 
\begin{eqnarray}
\frac{\partial u}{\partial \tau} &=& 
\frac{1}{n^2(\tau,R)} 
\left( \ex \dot{R}-\delta \sqrt{\dot{R}^2+n^2(\tau,R)} \right)
\label{tr12} \\
\frac{\partial u}{\partial \eta} &=& 
-\theta \left( \ex \dot{R}+\delta \sqrt{\dot{R}^2+n^2(\tau,R)} \right)^{-1}
\label{tr22} \\
\frac{\partial r}{\partial \tau} &=& 
\dot{R}
\label{tr32} \\
\frac{\partial r}{\partial \eta} &=& 
-\theta \delta \ex \sqrt{\dot{R}^2+n^2(\tau,R)}
\label{tr42} \end{eqnarray}
where the dot denotes a derivative with respect to proper time and 
the parameters $\ex$, $\theta$, $\delta$ can be taken independently
$\pm 1$.
We have also defined 
\be
n^2(\tau,R)=n^2(u(\tau,\eta=0),R)=
\frac{1}{12M^3}\Lx R^2
+k-
\frac{1}{6\pi^2M^3} \frac{\calm(\tau,R)}{R^3} 
\label{nnss} \ee
and denoted $\calm(\tau,R)\equiv \calm(u(\tau,\eta=0),R)$.

The values of the parameters $\ex$, $\theta$, $\delta$ are 
constrained by various physical considerations. It is convenient to
choose $\partial u/\partial \tau >0$. 
The opposite choice corresponds to
the reverse evolution, essentially exchanging energy ouflow with influx.
The requirement $\ex \calm_{,u} \geq 0$, imposed by the energy conditions,
indicates that we must associate $\ex=1$ with energy outflow (for which
$\dot{\calm}\equiv\partial \calm(\tau,R)/\partial \tau>0$) and $\ex=-1$ 
with energy influx (for which $\dot{\calm}<0$).
Finally we must demand that 
$\theta \delta \ex =1$, so that $\partial r/\partial \eta <0$ \cite{kraus}.
This requirement guarantees that the brane energy density is positive
\cite{binetruy,brax,tet1}. 
For an expanding brane, 
outside the horizon of the bulk metric ($n^2>0$), 
the above requirements are satisfied for $\delta=-1$, $\theta=-\ex$. 

It is straightforward to express the bulk energy-momentum tensor 
at the location of the brane in the coordinate system 
$({\tau},\eta)$. We find
\begin{eqnarray}
T^0_{~0}&=&\Lx
-\frac{1}{2\pi^2} \frac{\calm'}{R^3}
+\frac{1}{\dot{R}-\ex\sqrt{\dot{R}^2+n^2}} \frac{1}{2\pi^2} 
\frac{\dot{\calm}}{R^3}
\label{t00e2} \\
T^1_{~1}=T^2_{~2}=T^3_{~3}
&=& \Lx -\frac{1}{6\pi^2} \frac{\calm''}{R^2}
\label{t11e2} \\
T^4_{~4}&=&
\Lx
-\frac{1}{2\pi^2} \frac{\calm'}{R^3}
-\frac{1}{\dot{R}-\ex\sqrt{\dot{R}^2+n^2}} \frac{1}{2\pi^2} 
\frac{\dot{\calm}}{R^3}
\label{t44e2} \\
T^0_{~4}
&=&\frac{1}{-\ex\dot{R}+\sqrt{\dot{R}^2+n^2}} \frac{1}{2\pi^2} 
\frac{\dot{\calm}}{R^3},
\label{t04e2} 
\end{eqnarray}
where we have made use of $\delta=-1$, $\theta=-\ex$. 
A dot denotes a partial derivative with respect to proper time $\tau$, while
a prime a partial derivative with respect to $R$.

The equations governing the cosmological evolution on the brane 
can be obtained as before.
We assume that the brane energy-momentum tensor has the form
of eq. (\ref{branet}).
The evolution on the brane is governed by the equations 
(\ref{energ}), (\ref{chi}), (\ref{phi}) and
\be
H^2=\left(\frac{\dot{R}}{R}\right)^2=
\frac{1}{144 M^6}\left(\rht^2+2V\rht\right)
+\chi+\phi+\lx,
\label{last} \ee
with $\lx=(V^2/12M^3-\Lx)/12M^3$.

For the particular form of the energy-momentum tensor that we are considering,
it is possible to show that 
\be
\chi+\phi=\frac{1}{6\pi^2M^3}\frac{\calm(\tau,R)}{R^4}
\label{chiphi2} \ee
satisfies eqs. (\ref{chi}), (\ref{phi}).
This permits us to write eqs. (\ref{last}), (\ref{energ})
in the form
\begin{eqnarray}
H^2
&=&\frac{1}{144 M^6}\rht^2+ \frac{1}{6\mpl^2}\rht
+\frac{1}{6\pi^2M^3}\frac{\calm}{R^4}
-\frac{k}{R^2}
+\lx
\label{hubble22} \\
\dot{\rht}+3 H (\rht+\pt)&=&
-\frac{12M^3}{\pi^2V}\frac{\dot{\calm}}{R^4}
\frac{1}{1-\ex\frac{12M^3H}{V}+\frac{\rht}{V}} 
\label{energ2}
\end{eqnarray} 
with $\mpl^2=12M^6/V$.

An alternative way to derive eq. (\ref{hubble22}) is by considering
the junction conditions for the boundary defined by the brane motion. 
This procedure has been followed in refs. \cite{vaidya,chamblin} for $\ex=1$,
with a result in
agreement with eq. (\ref{hubble22}).

\subsection{The cosmological evolution}

The cosmological evolution on the brane is described by 
eqs. (\ref{hubble22}), (\ref{energ2}). The 
function $\calm(\tau,R)$ is required by the various energy conditions
to satisfy
$\ex \dot{\calm} \geq 0$, 
$\calm' \geq 0$, 
$\calm'' \leq 0$, 
$\calm' \geq - R \calm''/3$.
We recognize the effective cosmological constant and curvature terms,
as well as the linear and quadratic contributions from the brane
energy density. There is also a generalization of the ``mirage'' radiation
term with $\calm\to\calm(\tau,R)$.

These equations with $\ex=1$ have been used in the study of energy outflow from
the brane through the emission of bulk gravitons \cite{hebecker,vaidya}. 
The cross-section for bulk graviton 
production in collisions of brane particles was calculated and the
energy loss was determined using the Boltzmann equation.
An expression for
$\dot{\calm}$ was obtained by identifying the r.h.s. of eq. (\ref{energ2})
with the computed energy loss, under the assumption $\calm=\calm(\tau)$. 
It is possible, however, that the form of the bulk metric has a non-trivial
dependence on $R$, associated with the presence of a bulk fluid. 
It is also possible that energy can be deposited on the brane, 
so that the choice $\ex=-1$ must be made. 

In the low-energy regime the last factor in the r.h.s. of eq. (\ref{energ2})
becomes 1, so that the same equations describe energy outflow or influx,
depending on the sign of $\dot{\calm}$.
For $\calm=\calm(\tau)$ 
we can put eqs. (\ref{hubble22}), (\ref{energ2}) in the form
\begin{eqnarray}
H^2 &=&
\frac{1}{6\mpl^2}(\rht+\rht_r)
-\frac{k}{R^2}
+\lx
\label{hubble223} \\
\dot{\rht}+3 H (\rht+\pt)&=&-(\dot{\rht}_r+4 H \rht_r),
\label{energ23}
\end{eqnarray}
with 
$\rht_r=12M^3\calm(\tau)/(\pi^2 V R^4)$.
These equations describe an expanding Universe in which 
brane matter can be transformed to ``mirage'' radiation, or the opposite,
while the total energy is conserved. 
In the high-energy regime the two possible values of $\ex$ result in 
different forms of eq. (\ref{energ2}).

We can also allow for a non-trivial dependence of $\calm$ on $R$.
As a simple example, we can write
$\calm=\zeta \, \gamma(\tau) R^n$ with $\zeta=\pm 1$, $n$ integer, and
$\gamma(\tau)$ a positive-definite function.
The various energy conditions constrain the possible values of 
$n$ and $\zeta$. Apart from the case $(n,\zeta)=(1,1)$ that we discussed
above, we have the possibilites $(n,\zeta)=(1,1),(-1,-1),(-2,-1)$.

In the low-energy regime, the first case results in 
\begin{eqnarray}
H^2 &=&
\frac{1}{6\mpl^2}(\rht+\rht_m)
-\frac{k}{R^2}
+\lx
\label{hubble224} \\
\dot{\rht}+3 H (\rht+\pt)&=&-(\dot{\rht}_m+3 H \rht_m),
\label{energ24}
\end{eqnarray}
with 
$\rht_m=12M^3\gamma(\tau)/(\pi^2 V R^3)$.
These equations describe the transformation of brane matter to ``mirage'' 
dust, or
the opposite, in an expanding Universe. 
The bulk fluid that is associated with the necessary energy-momentum
tensor for such a solution
may seem exotic, but can have a straightforward physical interpretation.
As an example, let us assume that $\gamma$ is constant so that there is
no energy exchange between the brane and the bulk ($T^0_{~4}=0$).
In both coordinate systems $(u,r)$ and $(\tau,\eta)$
the bulk contains, apart from the negative cosmological constant, 
an anisotropic perfect fluid 
with $T^A_{~B}={\rm diag}(-\rho,p,p,p,{\rm p})$ and
$p=0$, ${\rm p}=-\rho$.
An analogous energy-momentum tensor in four dimensions is related to 
the gravitational field of a monopole \cite{wang,vilenkin}.

The other two possibilities with $(n,\zeta)=(-1,-1),(-2,-1)$
result in a negative contribution to the r.h.s. of eq. (\ref{hubble22}).
For constant $\gamma$ a linear combination of 
the second one and a component with $(n,\zeta)=(1,1)$ corresponds to a 
Reissner-Nordstrom-AdS bulk space. In both cases the presence of the
bulk influences substantially only 
the early-time cosmological evolution on the brane (small $R$).

\section{Conclusions}

The main purpose of this work has been to derive exact solutions of
the Einstein equations for the bulk-brane system in order to explore the
new physical behaviour that could possibly appear. 
The emphasis has been put on making appropriate 
assumptions for the form of the bulk and brane matter 
that could lead to an exact solution. This paper is a continuation of
ref. \cite{pantelis}, allowing for a time-dependent bulk within which
the brane motion takes place.

In the first model that we discussed the bulk energy is distributed
over the whole range of a static fourth spatial dimension. It 
induces an expansion along the other three spatial dimensions.
When a brane is embedded in this background, the cosmological
evolution seen by a brane observer results from a combination of the 
expansion of the background and the expansion induced through the
brane motion. There is also energy exchange between the brane and the
bulk. An expanding brane Universe is consistent only with energy outflow,
while a contracting one with influx. 
The effective Friedmann equation for the brane expansion includes  
terms proportional to the bulk and the brane energy density. 
During an era of
expansion the brane energy density is diluted because of the increase
of the scale factor and the outflow. As a result, after a certain time
the rate of expansion is completely dominated by the bulk energy density. 
This model also allows for periods of accelerated expansion.
These are short and are expected to be followed by periods of contraction.

In the second model we considered a bulk with a generalized Vaidya metric
\cite{wang}, describing energy flow along the fourth spatial dimension.
A particular form of this metric has been employed in the past 
for the discussion of energy outflow from the brane through the 
emission of bulk gravitons \cite{hebecker,vaidya}.
When the brane is embedded in the generalized Vaidya bulk,
the effective Friedmann equation for the expansion seen by a
brane observer includes a contribution from a ``mirage'' matter component.
Moreover, the brane energy density can be transformed to ``mirage''
energy density, or the opposite. The ``mirage'' component can have
various effective equations of state. For a particular bulk configuration
it  behaves like ``mirage'' dust, i.e. it is pressureless.

The overall picture that emerges from the models we studied here and in
ref. \cite{pantelis} indicates that the effects of the bulk on the
cosmological evolution on the brane can go beyond the presence of a
``mirage'' radiation component. The bulk energy density can drive the expansion
even for very dilute brane matter. Also the 
``mirage'' effects can have several forms, induced
by the brane motion within various possible backgrounds:
Schwarzschild-AdS, Vaidya-AdS, Reissner-Nordstrom-AdS, 
star-AdS \cite{pantelis}, various anisotropic fluids in AdS
etc. Energy exchange between the bulk and the brane is a common occurence. 
Generically
the cosmological evolution on the brane for low energy density 
is that of a Friedmann-Robertson-Walker Universe with several novel features.

\vskip 0.5cm
\centerline{\bf\large Acknowledgments}
\vskip .3cm

I would like to thank P. Apostolopoulos, T. Christodoulakis,
E. Leeper and R. Maartens for many useful discussions.
The work of N. Tetradis was supported through the RTN contract
HPRN--CT--2000--00148 of the European Union and the research
program ``Pythagoras'', grant number 016.


\vskip 1.5cm


\begin{thebibliography}{999}

\bibitem{rs} 
L. Randall and R. Sundrum, 
Phys. Rev. Lett. {\bf 83} (1999) 3370 
[arXiv:hep-th/9905221];
Phys. Rev. Lett. {\bf 83} (1999) 4690 
[arXiv:hep-th/9906064]. 

\bibitem{binetruy}
P.~Binetruy, C.~Deffayet and D.~Langlois,
Nucl.\ Phys.\ B {\bf 565} (2000) 269
[arXiv:hep-th/9905012];
\\
P.~Binetruy, C.~Deffayet, U.~Ellwanger and D.~Langlois,
Phys.\ Lett.\ B {\bf 477} (2000) 285
[arXiv:hep-th/9910219].

\bibitem{csaki}
C.~Csaki, M.~Graesser, C.~F.~Kolda and J.~Terning,
Phys.\ Lett.\ B {\bf 462} (1999) 34
[arXiv:hep-ph/9906513];
\\
J.~M.~Cline, C.~Grojean and G.~Servant,
Phys.\ Rev.\ Lett.\  {\bf 83} (1999) 4245
[arXiv:hep-ph/9906523].

\bibitem{kraus}
P.~Kraus,
JHEP {\bf 9912} (1999) 011
[arXiv:hep-th/9910149].

\bibitem{brax}
P.~Brax and C.~van de Bruck,
Class.\ Quant.\ Grav.\  {\bf 20} (2003) R201
[arXiv:hep-th/0303095];
\\
R.~Maartens,
arXiv:gr-qc/0312059.

\bibitem{mirage}
A.~Kehagias and E.~Kiritsis,
JHEP {\bf 9911} (1999) 022
[arXiv:hep-th/9910174].
 
\bibitem{hebecker} 
A.~Hebecker and J.~March-Russell,
Nucl.\ Phys.\ B {\bf 608} (2001) 375
[arXiv:hep-ph/0103214].

\bibitem{tet1}
E.~Kiritsis, G.~Kofinas, N.~Tetradis, T.~N.~Tomaras and V.~Zarikas,
JHEP {\bf 0302} (2003) 035
[arXiv:hep-th/0207060].

\bibitem{vandebruck}
C.~van de Bruck, M.~Dorca, C.~J.~A.~Martins and M.~Parry,
Phys.\ Lett.\ B {\bf 495} (2000) 183
[arXiv:hep-th/0009056].

\bibitem{tet2}
N.~Tetradis,
Phys.\ Lett.\ B {\bf 569} (2003) 1
[arXiv:hep-th/0211200].

\bibitem{dilaton}
H.~A.~Chamblin and H.~S.~Reall,
Nucl.\ Phys.\ B {\bf 562} (1999) 133
[arXiv:hep-th/9903225];
\\
C.~Barcelo and M.~Visser,
Phys.\ Rev.\ D {\bf 63} (2001) 024004
[arXiv:gr-qc/0008008];
\\
K.~i.~Maeda and D.~Wands,
Phys.\ Rev.\ D {\bf 62} (2000) 124009
[arXiv:hep-th/0008188];
\\
A.~Mennim and R.~A.~Battye,
Class.\ Quant.\ Grav.\  {\bf 18} (2001) 2171
[arXiv:hep-th/0008192].

\bibitem{vaidya}
D.~Langlois, L.~Sorbo and M.~Rodriguez-Martinez,
Phys.\ Rev.\ Lett.\  {\bf 89} (2002) 171301
[arXiv:hep-th/0206146];
\\
E.~Leeper, R.~Maartens and C.~F.~Sopuerta,
Class.\ Quant.\ Grav.\  {\bf 21} (2004) 1125
[arXiv:gr-qc/0309080].

\bibitem{chamblin}
A.~Chamblin, A.~Karch and A.~Nayeri,
Phys.\ Lett.\ B {\bf 509} (2001) 163
[arXiv:hep-th/0007060]. 

\bibitem{rothstein}
I.~Z.~Rothstein,
Phys.\ Rev.\ D {\bf 64} (2001) 084024
[arXiv:hep-th/0106022].

\bibitem{birm}
D.~Birmingham,
Class.\ Quant.\ Grav.\  {\bf 16} (1999) 1197
[arXiv:hep-th/9808032].

\bibitem{gregory}
S.~Mukohyama, T.~Shiromizu and K.~i.~Maeda,
Phys.\ Rev.\ D {\bf 62} (2000) 024028
[Erratum-ibid.\ D {\bf 63} (2001) 029901]
[arXiv:hep-th/9912287];
\\
P.~Bowcock, C.~Charmousis and R.~Gregory,
Class.\ Quant.\ Grav.\  {\bf 17} (2000) 4745
[arXiv:hep-th/0007177].

\bibitem{pantelis}
P.~S.~Apostolopoulos and N.~Tetradis,
arXiv:hep-th/0404105.

\bibitem{israel}
W.~Israel,
Nuovo Cim.\ B {\bf 44} (1966) 1.

\bibitem{wang}
A.~z.~Wang and Y.~m.~Wu,
Gen.\ Relativ.\ Grav. {\bf 31} (1999) 107
[arXiv:gr-qc/0207121].

\bibitem{vilenkin}
M.~Barriola and A.~Vilenkin,
Phys.\ Rev.\ Lett.\  {\bf 63} (1989) 341.





\end{thebibliography}
\end{document}